\documentclass[aps,prl,reprint,superscriptaddress]{revtex4-1}
\usepackage{mathrsfs}
\usepackage{amsmath,gensymb}
\usepackage{amsfonts}
\usepackage{amssymb}
\usepackage{amsthm}
\usepackage{graphicx}
\usepackage{natbib}
\usepackage{color}
\usepackage{hyperref}
\usepackage{bm}
\usepackage[caption=false]{subfig}
\usepackage{verbatim}
\RequirePackage{lineno}
\providecommand{\U}[1]{\protect\rule{.1in}{.1in}}

\begin{document}
\title{Emergence of Unconventional Interfacial Magnetic Phenomenon in  \\ Topological Insulator-Based Magnetic Heterostructures}
\author{Dhavala Suri}
\email{dhavala.suri@gmail.com}
\affiliation{Tata Institute of Fundamental Research, Hyderabad, Telangana 500046, India}
\author{Archit Bhardwaj}
\affiliation{Tata Institute of Fundamental Research, Hyderabad, Telangana 500046, India}
\author{Satyaki Sasmal}
\affiliation{Tata Institute of Fundamental Research, Hyderabad, Telangana 500046, India}
\author{Karthik V. Raman}
\email{kvraman@tifrh.res.in}
\affiliation{Tata Institute of Fundamental Research, Hyderabad, Telangana 500046, India}

\begin{abstract} 
In a topological insulator (TI)/magnetic insulator (MI) hetero-structure, large spin-orbit coupling of the TI and inversion symmetry breaking at the interface could foster  non-planar spin textures such as skyrmions at the interface. This is observed as topological Hall effect in a conventional Hall set-up. While this effect has been observed at the interface of TI/MI, where MI beholds perpendicular magnetic anisotropy, electronic transport features sensitive to the interface of MI with in-plane magnetic anisotropy is evidently under-reported. In this work, we study Bi$_2$Te$_3$/EuS hetero-structure using planar Hall measurements and observe planar topological Hall-like features. Additionally, a spontaneous planar Hall signal is observed which approaches a maximum value when the current and magnetic field directions are aligned perpendicularly within the sample plane. This response maybe attributed to the underlying planar anisotropy of the interfacial magnetic state; thereby demonstrating the importance of PHE for sensitive detection and characterization of non-trivial magnetic phase that has evaded exploration in the TI/MI interface studies.
\end{abstract}

\maketitle

Spin textures with non-trivial topology, also known as skyrmions are commonly observed in materials where  non-centrosymmetry of the crystal leads to an unconventional exchange  arising from the Dzyaloshinskii-Moriya (DM) interaction \cite{Muhlbauer915,doi:10.1063/1.5130423}. Such spin textures can also be stabilized in proximity coupled hetero-structures of heavy metal (HM) and ferromagnet, by virtue of large spin-orbit coupling (SOC) of the HM and inversion symmetry breaking at the interface \cite{Shao2019}. Topological insulators (TIs) offer yet another robust platform in this regard, whose high SOC is proven to give rise to skyrmions in proximitized ferromagnetic metals and insulators  \cite{Yasuda2016,whao,PhysRevLett.121.096802,PhysRevLett.119.176809,Zhang2018,Chen2019,Li2021}. However, in comparison with ferromagnetic metals, manipulation of skyrmions in magnetic insulators (MI) provide a technological advantage due to the lower dissipative losses. Hence, the recent report of skyrmions appearing at the interface between an MI and a TI  marks a significant progress  \cite{Li2021}. Furthermore, investigating these TI/MI interfaces is fundamentally crucial to understand the interplay between the real-space topology of spin-textures and the band topology of TI \cite{PhysRevB.98.060401}.

The TI/MI interface has been extensively investigated for promising emergent quantum states \cite{Che2018,Lang2014,Tange1700307,Fanchiang2018,Liang2015,Katmis2016,PhysRevLett.110.186807,PhysRevLett.119.027201,PhysRevB.99.064423} that arise by virtue of time reversal symmetry breaking of the TI surface states. However, the perspective of inverse proximity effect (IPE) whereby the large SOC of the TI affecting the magnetic properties of MI at the interface has received limited attention. Here we present the development of unconventional magnetism in EuS -- an MI, interfaced with the Bi$_2$Te$_3$, a three dimensional TI. EuS is a Heisenberg ferromagnetic insulator with a Curie temperature of 16.6~K. In contrast to the  experiment by Li et. al. \cite{Li2021}, that used an MI with perpendicular magnetic anisotropy (PMA), we study EuS  which has an in-plane anisotropy \cite{Wei2016}. Studies of TI/EuS have shown strong enhancement in the Curie temperature of interfacial EuS with a weak out-of-plane anisotropy \cite{Katmis2016}, which was later found to be influenced by carrier mediated Ruderman–Kittel–Kasuya–Yosida interactions \cite{PhysRevLett.119.027201,Mathimalar2020}. Theoretically, formation of skyrmion states in such in-plane anisotropy materials emerge due to the development of weak out-of-plane anisotropy \cite{PhysRevApplied.12.064054,PhysRevB.96.201301}. Although spin canting of EuS moments at the interface was observed in these studies, the nature of spin texture  remains an open question. In this work we study TI/EuS electronically, and find features that are signatures of plausible non-trivial spin textures \cite{PhysRevApplied.12.064054,Du2014,PhysRevB.98.094404} at the interface of TI/MI.  Interestingly, their electronic signature was not apparent in conventional Hall measurements, requiring a planar Hall measurement configuration to unveil its presence.  

Planar Hall effect (PHE) refers to the development of a transverse voltage in response to longitudinal current and an in-plane magnetic field H, such that the influence of Lorentz force is absent \cite{Taskin2017,bhardwaj2021observation}. In a conventional ferromagnet or a TI, PHE depends on $\varphi$, the in-plane angle between current and magnetic field, as R$_{xy} \propto \sin 2\varphi$ and R$_{xx} \propto \cos^2 \varphi$, where R$_{xy}$ is the planar Hall resistance (PHR) \cite{doi:10.7566/JPSJ.84.104708,Li2021} and  R$_{xx}$ is the planar magneto-resistance (PMR). As a result, the R$_{xx}$ and R$_{xy}$ signals are  $\pi/4$ out-of-phase in $\varphi$. In addition, since the Lorentz force is absent in this configuration, the R$_{xx}$ and R$_{xy}$ signals are purely symmetric with respect to H. Therefore, any deviation from the above  is  indicative of additional mechanisms that drive the unconventional PHE. In our study on Bi$_2$Te$_3$/EuS interface, we provide a comprehensive picture of such unconventional responses appearing in the PHR and PMR signals at low magnetic fields and associate them to the development of unconventional magnetism in the interfacial EuS of the TI/MI hetero-structure.

\par In this work, epitaxial films of Bi$_2$Te$_3$~(14 quintuple layers) are grown on Al$_2$O$_3$~(0001) substrate, followed by EuS (5 nm)/AlO$_x$ (x$\sim{1.5-2}$), using molecular beam epitaxy technique (refer SI for details). The surface-state effect of TI is turned off by optimising the growth conditions such that the Fermi level is placed deep inside the conduction band. For the PHE studies, the as-grown films are patterned into a Hall bar device of channel length and width  $\approx 3$~mm and $\approx 1.5$~mm respectively, by mechanical scratching to eliminate any possible contamination due to lithographic processes. 
\begin{figure}[tbh]
\centering
\includegraphics[width=8cm]{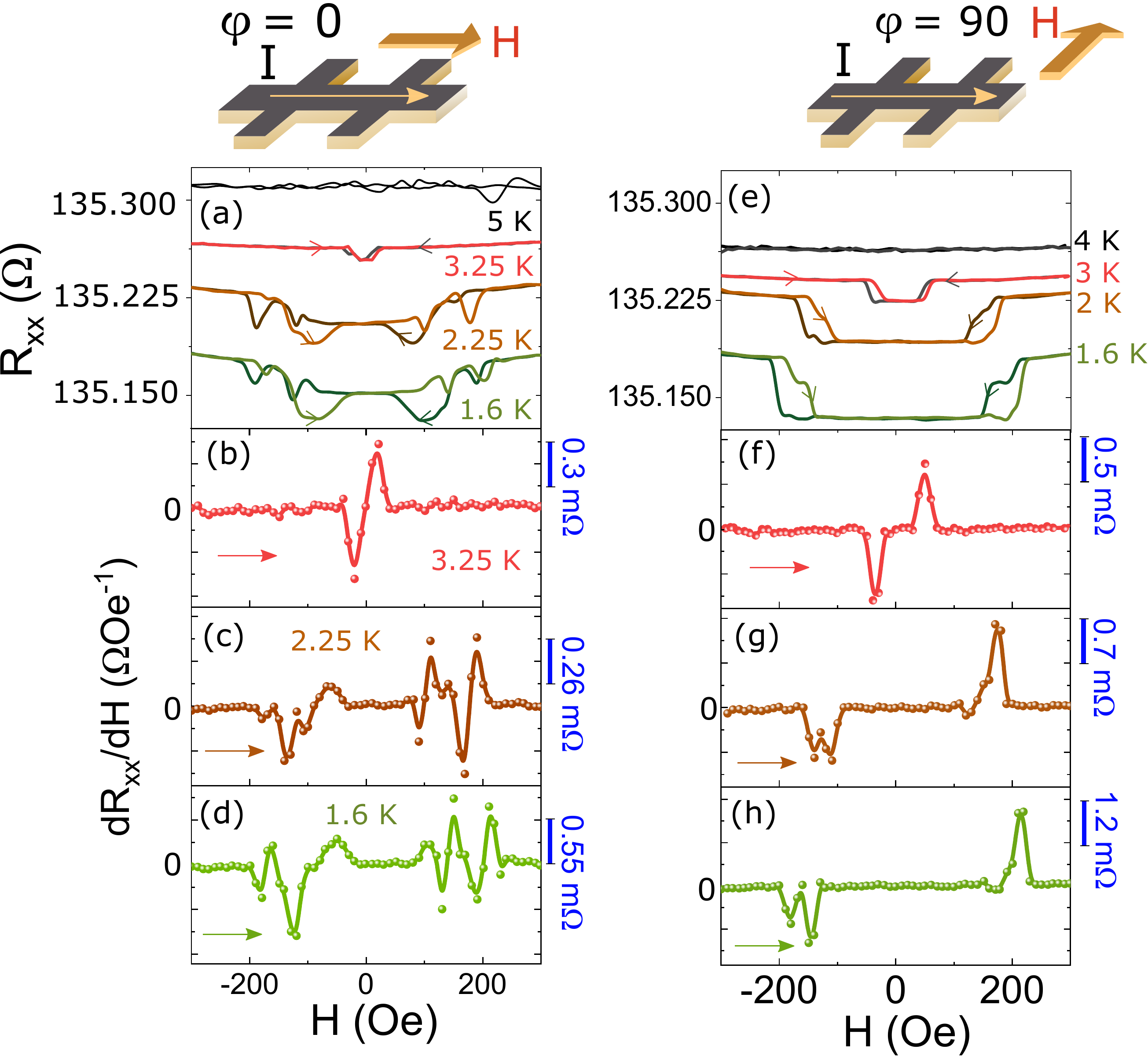}
\caption{PMR (R$_{xx}$) versus in-plane magnetic field (scan direction shown by arrows) at different temperatures, in the configuration (a) $\varphi = 0^{\circ}$ (e)  $\varphi = 90^{\circ}$.  The corresponding first derivatives of only the forward scans, in the configurations (b), (c) and (d) $\varphi = 0^{\circ}$ (f), (g) and (h) $\varphi = 90^{\circ}$.}
\label{Rxx}
\end{figure}

\par In fig.~\ref{Rxx}, we investigate the PMR of the patterned device in low magnetic fields ($\lvert H \rvert <$ 300~Oe) within the coercive field. The alignment of film-plane parallel to magnetic field is carefully performed by monitoring the minimum of the Hall voltage with varying out-of-plane angle $\theta$ (refer to SI). Fig.~\ref{Rxx} (a) and (e) shows R$_{xx}$ vs H, as a function of temperature (T) in the $\varphi = 0^{\circ}$ and $\varphi = 90^{\circ}$ configuration respectively, where $\varphi$ is the angle between current and in-plane magnetic field. We find that R$_{xx}$ is nearly independent of magnetic field down to 5~K. However, for T $<$ 4~K, a step-like feature develops at low fields with the emergence of two hysteresis loops symmetric about H = 0. The origin of this hysteresis is different from the conventional hysteresis of a ferromagnet. In a ferromagnet, the resistance change occurs close to the film's coercive field upon reversing the direction of H. However in these devices, the transitions are observed much before the field reversal. Hence, the observed unconventional step-like response in PMR indicates an emergent magnetic phase transition as seen by the derivative curves in fig.~\ref{Rxx} (b)-(d) and (f)-(h) (shown only for the forward scans).  On lowering the temperature, the magnetic field corresponding to these phase transition increases, implying higher stability of the emergent magnetic phase. Similar phase transitions have been commonly observed in systems such as MnSi that hosts skyrmionic spin textures \cite{Liang2015,Du2014,https://doi.org/10.1002/adfm.202008521,Yokouchieaat1115,PhysRevB.87.134424}; this hints to the possible emergence of robust non-trivial spin textures at the Bi$_2$Te$_3$/EuS interface in our devices. Magnetometry studies on these films (see SI) reveal that these transitions appear in the non-saturating state of EuS when the competing effects of internal exchange and anisotropy fields form a stable magnetic phase. Together with the  inversion symmetry breaking and the large SOC of Bi$_2$Te$_3$ at the Bi$_2$Te$_3$/EuS interface, conditions become conducive to generate strong interfacial DM interactions favoring the formation of non-trivial spin-textures. The phase transition in $\varphi = 0^{\circ}$ configuration is accompanied by oscillations [fig.~\ref{Rxx} (c), (d)], which maybe attributed to relatively larger spin fluctuations in this geometry as compared with $\varphi = 90^{\circ}$ [fig.~\ref{Rxx} (g), (h)]. Interestingly, the PMR curves in $\varphi = 0^{\circ}$ and $\varphi = 90^{\circ}$ configuration exhibit contrasting responses; with the latter configuration showing a larger region of plateau at low field and a relatively sharper transition as interpreted from the first derivative plots. All these observations suggest an in-plane anisotropy in probing these spin-textures. 

\begin{figure*}[th]
\centering
\includegraphics[width=\textwidth]{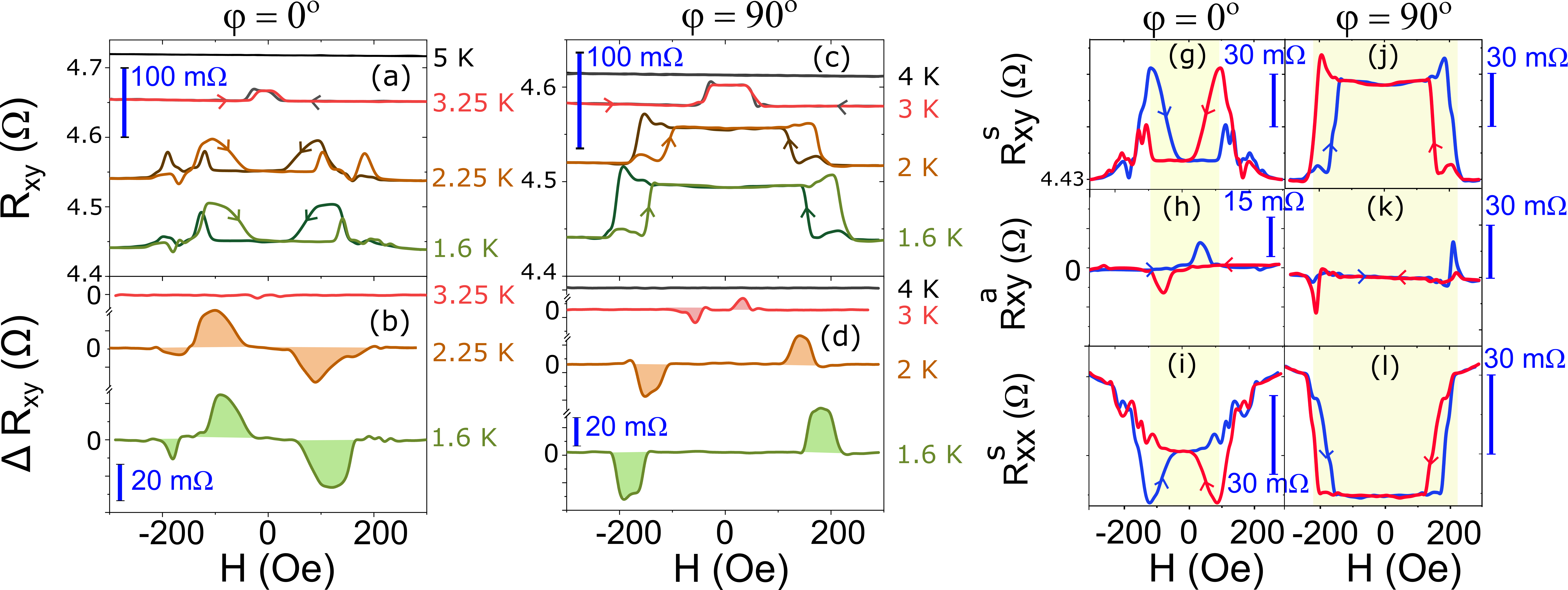}
\caption{(a), (c) Transverse resistance R$_{xy}$ versus in-plane magnetic field H at different temperatures. (b), (d) The difference curves of forward and backward moving curves, representing hysteresis. (g), (j) Symmetric component of R$_{xy}$,  R$_{xy}^s$ vs H; (h), (k) the anti-symmetric component R$_{xy}^a$ vs H; (i), (j) the symmetric component  R$_{xx}$ vs H, all at T = 1.6~K. Measurement configurations of the curves ($\varphi = 0^{\circ}$, $\varphi = 90^{\circ}$), are as mentioned on top. Scan direction of  magnetic field is shown by arrows on the curves. R$_{xy}^s$ = $\frac{R^{+H}_{xy}+ R^{-H}_{xy}}{2}$, ~R$_{xy}^a = \frac{R^{+H}_{xy} - R^{-H}_{xy}}{2}$, 
 R$_{xx}^s = \frac{R^{+H}_{xx} + R^{-H}_{xx}}{2}$.}
\label{Rxy}
\end{figure*}

\par Further, we analyse the PHR signal (R$_{xy}$) which was measured simultaneously with R$_{xx}$  for the above two measurement configurations of $\varphi$. Fig.~\ref{Rxy} (a) and (c) show R$_{xy}$ vs H measured at different temperatures in $\varphi = 0^{\circ}$ and $\varphi = 90^{\circ}$ configurations, respectively. We notice a similar step-like profile at low fields below 4~K with the same form of anisotropy and hysteresis as in the R$_{xx}$ signal, however predominantly with an inverted signal i.e., a drop in R$_{xx}$ corresponds to a rise in R$_{xy}$. An offset in the magnitude of  R$_{xy}$ corresponds to minor contribution from R$_{xx}$, which can be subtracted as constant baseline (see SI).  To characterize the magnetic phase formed at the Bi$_2$Te$_3$/EuS interface, we analyse the emergence of hysteresis in R$_{xy}$  by performing an elementary subtraction of the forward and backward scans ($\Delta$R$_{xy}$) in fig.~\ref{Rxy} (b) and (d). At 1.6~K, $\Delta$R$_{xy}$ peaks at $\approx$ 80~Oe and $\approx$ 160~Oe in the $\varphi = 0^{\circ}$ and $\varphi = 90^{\circ}$ configurations, respectively. Here, the sign reversal and the broadening of the hysteresis zone with cool down indicate the enhanced stabilization of the particular magnetic phase \cite{LIU2020268}. Similar responses in PHR accompanied by a hysteresis has been reported in studies of MnSi and FeGe \cite{doi:10.7566/JPSJ.84.104708,https://doi.org/10.1002/adfm.202008521} and attributed to the formation of topologically conserved spin-textures such as skyrmions. 

\par To examine the formation of spin-textures at the Bi$_2$Te$_3$/EuS interface using PHE studies by decomposing the PHR and PMR signal into two components -- the symmetric signal (R$_{xy}^s$ and R$_{xx}^s$) of a pure planar Hall response and the anti-symmetric signal (R$_{xy}^a$ and R$_{xx}^a$) arising from an anomalous PHE contribution. Fig.~\ref{Rxy} shows R$_{xy}^s$, R$_{xy}^a$  and R$_{xx}^s$ versus H plotted for the $\varphi = 0^{\circ}$ [fig.~\ref{Rxy}(g), (h) and (i)] and $\varphi = 90^{\circ}$ [fig.~\ref{Rxy} (j), (k) (l)] configuration at T = 1.6~K. The R$_{xy}^s$ and R$_{xx}^s$ show inverted behavior with a similar range of variation: $\approx$ 65 m$\Omega$. For simplicity, in the following discussions we focus our analysis only on the forward scan data. Similar arguments can be extended to the backward scan plots.  In the forward scan, R$_{xy}^s$ [fig~\ref{Rxy}~(g), (j)] and R$_{xx}^s$ [fig.~\ref{Rxy}~(i), (l)] exhibit a peak and drop respectively, at  H~$\approx$ -150~Oe (-200~Oe) in $\varphi = 0^{\circ}$ ($\varphi = 90^{\circ}$) configuration. They correspond to the magnetic phase transition described in fig.~\ref{Rxx}. Subsequently, a region of plateauing is observed in R$_{xy}^s$ and R$_{xx}^s$ indicating the stability of a magnetic phase. For the case of $\varphi = 90^{\circ}$ configuration, this plateau region persist over a broader range of H ($-~140$~Oe $\lesssim$  H  $\lesssim$  $~200$~Oe). Within this region, a  hump develops in R$_{xy}^a$ vs H, peaking at $\approx$ 50 Oe ($\approx$ 200~Oe) in $\varphi = 0^{\circ}$ ($\varphi = 90^{\circ}$) configuration  [fig~\ref{Rxy}~(h) and (k)], which resembles topological Hall effect (THE) \cite{Li2021}. We have verified that this signal arises from the non-trivial spin-texture at the interface captured in the true planar Hall measurement and not due to an out-of-plane component of H caused by a misalignment in the measurement (see SI).  Subsequent to the plateau region in R$_{xy}^s$ and R$_{xx}^s$, the signals oscillate over a small region in $\varphi = 0^{\circ}$ configuration between $\approx $ 100~Oe and 200~Oe after which the applied field aligns the interfacial spins along its direction. In contrast, the $\varphi = 90^\circ$ configuration shows an abrupt transition at 200~Oe. Very similar transitions but without any hysteresis or in-plane anisotropy in $\varphi$ was observed in Bi$_2$Se$_3$/EuS studies (see SI), owing  to unfavourable growth or interface conditions.  We further probe the presence of hump-like feature in  R$_{xy}^a$ by rotating the field out-of-plane. Here R$_{xy}^a$ vs H measured in $\theta = 30^{\circ}$ configuration show a weak hysteretic behavior at H $\approx$ 180~Oe resembling the THE signal (see SI). In this configuration, the hump appears at a similar value of H with respect to the PHE study, thereby ruling out the formation of a conventional out-of-plane skyrmion state. This is further supported by the disappearence of the hump-like feature in the out-of-plane $\theta = 90^{\circ}$configuration  (refer to SI), elucidating that the formation of non-trivial spin-textures preserves the in-plane magnetic anisotropy.  

\par
\begin{figure}[tbh]
\centering
\includegraphics[width=8cm]{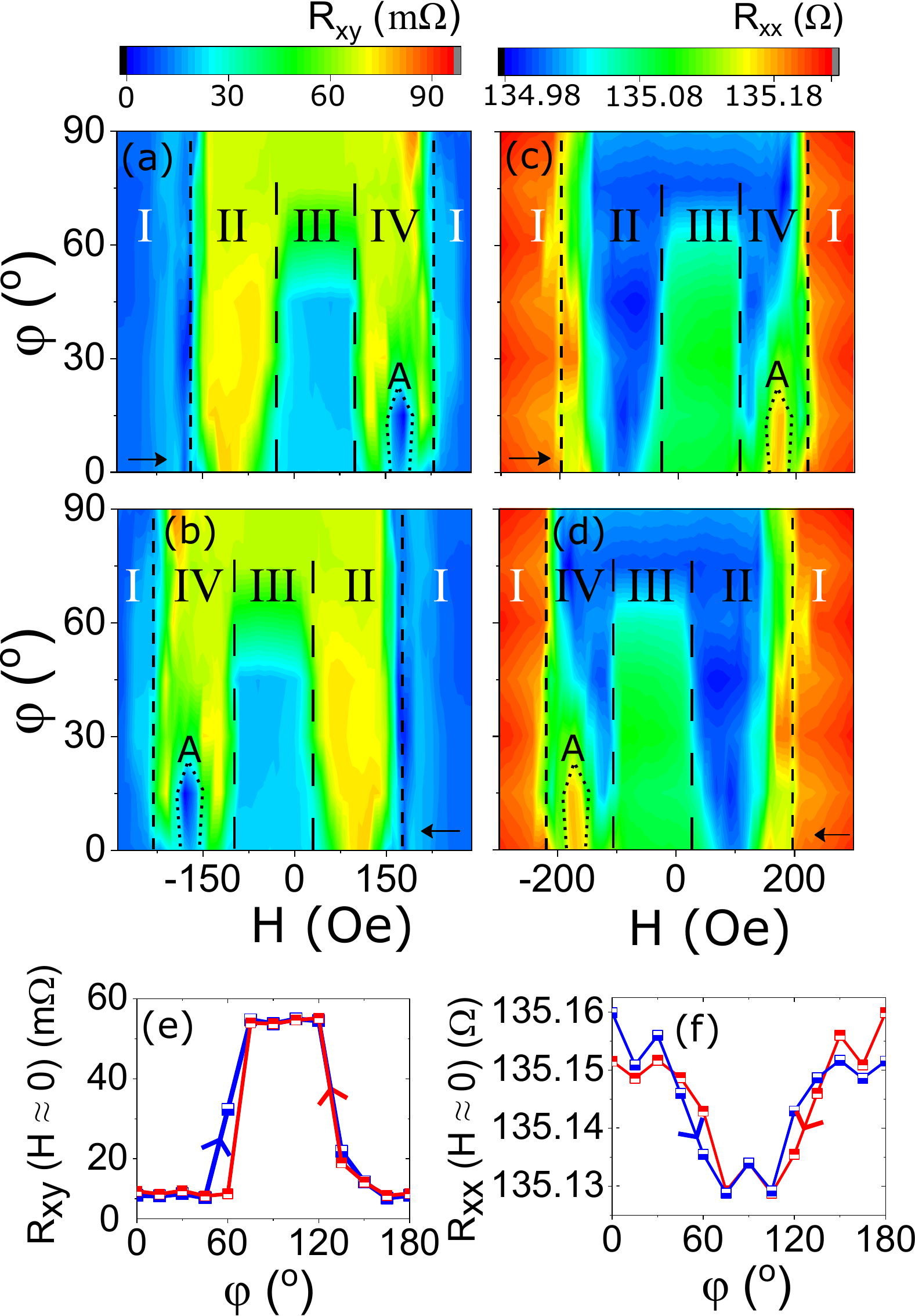}
\caption{Surface plots of symmetric components of  R$_{xy}^s$ vs H  (a) forward scan direction (b) backward scan direction. Surface plots of symmetric components of  R$_{xx}$ (c) forward scan direction (d) backward scan direction. Data is  symmetric about $\varphi = \pi/2$ (refer to SI for data of the full range $\varphi = 0^{\circ}$~to $180^{\circ}$). (e) R$_{xy}$ at H = 0  and (f) R$_{xx}$ at H = 0  extracted from the surface plot,  as a function of $\varphi$. A constant baseline of 4.43 $\Omega$ is subtracted from R$_{xy}$ for clarity.}
\label{3dplot}
\end{figure}

 We now describe the complete $\varphi$-dependence  ($0^{\circ} \leq \varphi \leq 90^{\circ}$) of PHE by plotting the surface maps of R$_{xy}^s$ and R$_{xx}^s$ in fig.~\ref{3dplot}, with varying H and $\varphi$.  Figure ~\ref{3dplot} (a), (c) and figure \ref{3dplot} (b), (d) shows the data corresponding to  forward and backward scans respectively, that appear as mirror images capturing all the critical features. Region I represents the magnetic field regime where collinear magnetization dominates, region II, III and IV represent the regime where non-collinear spin phases are stabilized. Interestingly, we find that the phase transitions of initial creation (region I $\rightarrow$ region II) and final annihilation (region IV $\rightarrow$ region I) of the interface spin-textures occur independent of $\varphi$. We also observe oscillations around the region A that persist only in the low $\varphi$ regime. This response may imply spin fluctuations concomitant with the phase transition. Furthermore, the  R$_{xy}^a$ vs H curve reveals a THE-like hump within the plateau region (see SI), corroborating the presence of non-trivial spin states. This plateau region broadens with increasing $\varphi$, shown by the merger of region II and III, indicating an anisotropy in the electronic detection of the spin-textures. 

\par It is worthwhile to remark that the conduction in the device is a result of electronic states at the interface and the bulk electronic transport in the TI. A pure TI without EuS in proximity shows no unconventional transport (see SI), implying that the features observed here, are dominated  solely from the interfacial magnetism. A clear deviation from the conventional PHE as observed at magnetic fields $>$ 1~T, also implies an unconventional mechanism at play (see SI). Furthermore, the observed anisotropy in the electronic detection can arise due to two possible scenarios. First, where the spin-texture has a planar asymmetry; here, the additional Berry phase acquired by the conduction electrons will depend on $\varphi$ and contribute to the anisotropy in the R$_{xy}^s$ and R$_{xx}^s$ signal. Second, assuming a symmetric spin texture, the anisotropy maybe attributed to deformation of the Fermi surface due to Zeeman effect on the Rashba states of the TI \cite{Soumyanarayanan2016}. However, this is unlikely owing to multiple crystal domain state of the films; even films that lack hexagonal symmetry of Bi$_2$Te$_3$ (verified by reflective high energy electron diffraction) demonstrate this anisotropy. Additionally, the spin-momentum locking property of the Dirac surface states of the TI is absent in our devices by design, since the Fermi level is tuned $\approx$ 0.35 eV (carrier density $\approx$ 10$^{20}$ cm$^{-3}$) deep into the conduction band from its edge.   

\par The presence of underlying planar asymmetry in the spin-textures is unveiled in the low-field measurement range within the plateau region (region III). This is shown in fig.~\ref{3dplot} (e)-(f) via a vertical line-cut in fig.~\ref{3dplot} (a), (b), (c) and (d) surface maps at H = 0. Interestingly, a spontaneous signal of PHR and PMR is observed that deviates from the conventional $\varphi$-dependence of the PHE signal. We observe the PHR and PMR signal to be symmetric about H = 0 and $\varphi= \pi/2$, with a periodicity of $\pi$. Such a symmetry may be correlated to the symmetry of the real-space spin textures at the interface. This observed anomalous signal is symmetric about H = 0, as seen in fig.~\ref{3dplot} (e), which is different from the proposed anomalous PHR in two dimensional systems \cite{PhysRevResearch.3.L012006,10.1088/1361-648X/abf783,PhysRevApplied.11.024064} or in case of Weyl  semimetals \cite{PhysRevB.103.214438}. Outside region III, there is no significant $\varphi$ dependence of R$_{xx}$ and  R$_{xy}$, conventional PHE signal of Bi$_2$Te$_3$ is observed at high fields (H $\geq$ 1 T  - see SI)  \cite{bhardwaj2021observation}.  

\par We remark that unlike the case of Bi$_2$Se$_3$/BaFeO \cite{Li2021} having a PMA MI, the magnetic response in our devices are observed only in the planar configuration. Theoretically, formation of skyrmions in a PMA material cannot simultaneously support textures of opposite topological charge \cite{PhysRevApplied.12.064054}. Hence, the PMA ensures the observation of THE at $\theta =$90$^\circ$ configuration associated to a specific sign of topological charge. In contrast, spin-textures in the in-plane anisotropy materials are observed to be different as they can simultaneously support textures of opposite topological charge. As a result, the net compensating spin textures become inaccessible in the out-of-plane measurement configuration ( $\theta =$ 90$^\circ$). It is likely that the interface of Bi$_2$Te$_3$/EuS support the formation of such in-plane magnetic states that show the $\pi/2$ symmetry in $\varphi$ \cite{PhysRevApplied.12.064054}. We however cannot rule out the possibility of other such spin-textures such as conical spirals or skyrmion tubes that are also proposed to give rise to planar THE \cite{PhysRevB.102.064430,PhysRevB.98.060401}. The inverted response of R$_{xx}$ and R$_{xy}$ with respect to H and $\varphi$, indicating a phase difference of $\pi$ between the two signals, also alludes to the unconventional origin consistent with studies on MnSi \cite{doi:10.7566/JPSJ.84.104708,https://doi.org/10.1002/adfm.202008521}. This requires a detailed theoretical model addressing the interface of TI/MI hetero-structures. Overcoming the challenge of imaging such interfacial spin-textures in the future may provide a means to decipher the precise nature of the spin textures., opening the path for further detailed experiment via spectroscopic techniques. 

\noindent To summarize, we have investigated the inverse proximity effect in a TI/MI system and reported the emergence of non-trivial interfacial phenomenon in the planar Hall configuration, via electronic transport studies. We find that the spontaneous planar Hall signal symmetric about $\varphi = \pi/2$, providing experimental evidence of unconventional magnetic correlations in the interfacial EuS. Absence of anomalous Hall and topological Hall effect in the out-of-plane magnetic field configuration, corroborates our understanding that the correlations are dominated by in-plane anisotropy and reveal the extreme sensitivity of the planar Hall measurement in detecting such interfacial effects. This work therefore calls for further experimental and theoretical investigations in probing interfacial spin textures.


\begin{thebibliography}{39}%
\makeatletter
\providecommand \@ifxundefined [1]{%
 \@ifx{#1\undefined}
}%
\providecommand \@ifnum [1]{%
 \ifnum #1\expandafter \@firstoftwo
 \else \expandafter \@secondoftwo
 \fi
}%
\providecommand \@ifx [1]{%
 \ifx #1\expandafter \@firstoftwo
 \else \expandafter \@secondoftwo
 \fi
}%
\providecommand \natexlab [1]{#1}%
\providecommand \enquote  [1]{``#1''}%
\providecommand \bibnamefont  [1]{#1}%
\providecommand \bibfnamefont [1]{#1}%
\providecommand \citenamefont [1]{#1}%
\providecommand \href@noop [0]{\@secondoftwo}%
\providecommand \href [0]{\begingroup \@sanitize@url \@href}%
\providecommand \@href[1]{\@@startlink{#1}\@@href}%
\providecommand \@@href[1]{\endgroup#1\@@endlink}%
\providecommand \@sanitize@url [0]{\catcode `\\12\catcode `\$12\catcode
  `\&12\catcode `\#12\catcode `\^12\catcode `\_12\catcode `\%12\relax}%
\providecommand \@@startlink[1]{}%
\providecommand \@@endlink[0]{}%
\providecommand \url  [0]{\begingroup\@sanitize@url \@url }%
\providecommand \@url [1]{\endgroup\@href {#1}{\urlprefix }}%
\providecommand \urlprefix  [0]{URL }%
\providecommand \Eprint [0]{\href }%
\providecommand \doibase [0]{http://dx.doi.org/}%
\providecommand \selectlanguage [0]{\@gobble}%
\providecommand \bibinfo  [0]{\@secondoftwo}%
\providecommand \bibfield  [0]{\@secondoftwo}%
\providecommand \translation [1]{[#1]}%
\providecommand \BibitemOpen [0]{}%
\providecommand \bibitemStop [0]{}%
\providecommand \bibitemNoStop [0]{.\EOS\space}%
\providecommand \EOS [0]{\spacefactor3000\relax}%
\providecommand \BibitemShut  [1]{\csname bibitem#1\endcsname}%
\let\auto@bib@innerbib\@empty
\bibitem [{\citenamefont {M{\"u}hlbauer}\ \emph {et~al.}(2009)\citenamefont
  {M{\"u}hlbauer}, \citenamefont {Binz}, \citenamefont {Jonietz}, \citenamefont
  {Pfleiderer}, \citenamefont {Rosch}, \citenamefont {Neubauer}, \citenamefont
  {Georgii},\ and\ \citenamefont {B{\"o}ni}}]{Muhlbauer915}%
  \BibitemOpen
  \bibfield  {author} {\bibinfo {author} {\bibfnamefont {S.}~\bibnamefont
  {M{\"u}hlbauer}}, \bibinfo {author} {\bibfnamefont {B.}~\bibnamefont {Binz}},
  \bibinfo {author} {\bibfnamefont {F.}~\bibnamefont {Jonietz}}, \bibinfo
  {author} {\bibfnamefont {C.}~\bibnamefont {Pfleiderer}}, \bibinfo {author}
  {\bibfnamefont {A.}~\bibnamefont {Rosch}}, \bibinfo {author} {\bibfnamefont
  {A.}~\bibnamefont {Neubauer}}, \bibinfo {author} {\bibfnamefont
  {R.}~\bibnamefont {Georgii}}, \ and\ \bibinfo {author} {\bibfnamefont
  {P.}~\bibnamefont {B{\"o}ni}},\ }\href@noop {} {\bibfield  {journal}
  {\bibinfo  {journal} {Science}\ }\textbf {\bibinfo {volume} {323}},\ \bibinfo
  {pages} {915} (\bibinfo {year} {2009})}\BibitemShut {NoStop}%
\bibitem [{\citenamefont {Mathur}\ \emph {et~al.}(2019)\citenamefont {Mathur},
  \citenamefont {Stolt},\ and\ \citenamefont {Jin}}]{doi:10.1063/1.5130423}%
  \BibitemOpen
  \bibfield  {author} {\bibinfo {author} {\bibfnamefont {N.}~\bibnamefont
  {Mathur}}, \bibinfo {author} {\bibfnamefont {M.~J.}\ \bibnamefont {Stolt}}, \
  and\ \bibinfo {author} {\bibfnamefont {S.}~\bibnamefont {Jin}},\ }\href
  {\doibase 10.1063/1.5130423} {\bibfield  {journal} {\bibinfo  {journal} {APL
  Materials}\ }\textbf {\bibinfo {volume} {7}},\ \bibinfo {pages} {120703}
  (\bibinfo {year} {2019})}\BibitemShut {NoStop}%
\bibitem [{\citenamefont {Shao}\ \emph {et~al.}(2019)\citenamefont {Shao},
  \citenamefont {Liu}, \citenamefont {Yu}, \citenamefont {Kim}, \citenamefont
  {Che}, \citenamefont {Tang}, \citenamefont {He}, \citenamefont {Tserkovnyak},
  \citenamefont {Shi},\ and\ \citenamefont {Wang}}]{Shao2019}%
  \BibitemOpen
  \bibfield  {author} {\bibinfo {author} {\bibfnamefont {Q.}~\bibnamefont
  {Shao}}, \bibinfo {author} {\bibfnamefont {Y.}~\bibnamefont {Liu}}, \bibinfo
  {author} {\bibfnamefont {G.}~\bibnamefont {Yu}}, \bibinfo {author}
  {\bibfnamefont {S.~K.}\ \bibnamefont {Kim}}, \bibinfo {author} {\bibfnamefont
  {X.}~\bibnamefont {Che}}, \bibinfo {author} {\bibfnamefont {C.}~\bibnamefont
  {Tang}}, \bibinfo {author} {\bibfnamefont {Q.~L.}\ \bibnamefont {He}},
  \bibinfo {author} {\bibfnamefont {Y.}~\bibnamefont {Tserkovnyak}}, \bibinfo
  {author} {\bibfnamefont {J.}~\bibnamefont {Shi}}, \ and\ \bibinfo {author}
  {\bibfnamefont {K.~L.}\ \bibnamefont {Wang}},\ }\href {\doibase
  10.1038/s41928-019-0246-x} {\bibfield  {journal} {\bibinfo  {journal} {Nature
  Electronics}\ }\textbf {\bibinfo {volume} {2}},\ \bibinfo {pages} {182}
  (\bibinfo {year} {2019})}\BibitemShut {NoStop}%
\bibitem [{\citenamefont {Yasuda}\ \emph {et~al.}(2016)\citenamefont {Yasuda},
  \citenamefont {Wakatsuki}, \citenamefont {Morimoto}, \citenamefont {Yoshimi},
  \citenamefont {Tsukazaki}, \citenamefont {Takahashi}, \citenamefont {Ezawa},
  \citenamefont {Kawasaki}, \citenamefont {Nagaosa},\ and\ \citenamefont
  {Tokura}}]{Yasuda2016}%
  \BibitemOpen
  \bibfield  {author} {\bibinfo {author} {\bibfnamefont {K.}~\bibnamefont
  {Yasuda}}, \bibinfo {author} {\bibfnamefont {R.}~\bibnamefont {Wakatsuki}},
  \bibinfo {author} {\bibfnamefont {T.}~\bibnamefont {Morimoto}}, \bibinfo
  {author} {\bibfnamefont {R.}~\bibnamefont {Yoshimi}}, \bibinfo {author}
  {\bibfnamefont {A.}~\bibnamefont {Tsukazaki}}, \bibinfo {author}
  {\bibfnamefont {K.~S.}\ \bibnamefont {Takahashi}}, \bibinfo {author}
  {\bibfnamefont {M.}~\bibnamefont {Ezawa}}, \bibinfo {author} {\bibfnamefont
  {M.}~\bibnamefont {Kawasaki}}, \bibinfo {author} {\bibfnamefont
  {N.}~\bibnamefont {Nagaosa}}, \ and\ \bibinfo {author} {\bibfnamefont
  {Y.}~\bibnamefont {Tokura}},\ }\href {\doibase 10.1038/nphys3671} {\bibfield
  {journal} {\bibinfo  {journal} {Nature Physics}\ }\textbf {\bibinfo {volume}
  {12}},\ \bibinfo {pages} {555} (\bibinfo {year} {2016})}\BibitemShut
  {NoStop}%
\bibitem [{\citenamefont {Wu}\ \emph {et~al.}(2020)\citenamefont {Wu},
  \citenamefont {Groß}, \citenamefont {Dai}, \citenamefont {Lujan},
  \citenamefont {Razavi}, \citenamefont {Zhang}, \citenamefont {Liu},
  \citenamefont {Sobotkiewich}, \citenamefont {Förster}, \citenamefont
  {Weigand}, \citenamefont {Schütz}, \citenamefont {Li}, \citenamefont
  {Gräfe},\ and\ \citenamefont {Wang}}]{whao}%
  \BibitemOpen
  \bibfield  {author} {\bibinfo {author} {\bibfnamefont {H.}~\bibnamefont
  {Wu}}, \bibinfo {author} {\bibfnamefont {F.}~\bibnamefont {Groß}}, \bibinfo
  {author} {\bibfnamefont {B.}~\bibnamefont {Dai}}, \bibinfo {author}
  {\bibfnamefont {D.}~\bibnamefont {Lujan}}, \bibinfo {author} {\bibfnamefont
  {S.~A.}\ \bibnamefont {Razavi}}, \bibinfo {author} {\bibfnamefont
  {P.}~\bibnamefont {Zhang}}, \bibinfo {author} {\bibfnamefont
  {Y.}~\bibnamefont {Liu}}, \bibinfo {author} {\bibfnamefont {K.}~\bibnamefont
  {Sobotkiewich}}, \bibinfo {author} {\bibfnamefont {J.}~\bibnamefont
  {Förster}}, \bibinfo {author} {\bibfnamefont {M.}~\bibnamefont {Weigand}},
  \bibinfo {author} {\bibfnamefont {G.}~\bibnamefont {Schütz}}, \bibinfo
  {author} {\bibfnamefont {X.}~\bibnamefont {Li}}, \bibinfo {author}
  {\bibfnamefont {J.}~\bibnamefont {Gräfe}}, \ and\ \bibinfo {author}
  {\bibfnamefont {K.~L.}\ \bibnamefont {Wang}},\ }\href@noop {} {\bibfield
  {journal} {\bibinfo  {journal} {Advanced Materials}\ }\textbf {\bibinfo
  {volume} {32}},\ \bibinfo {pages} {2003380} (\bibinfo {year}
  {2020})}\BibitemShut {NoStop}%
\bibitem [{\citenamefont {He}\ \emph {et~al.}(2018)\citenamefont {He},
  \citenamefont {Yin}, \citenamefont {Yu}, \citenamefont {Grutter},
  \citenamefont {Pan}, \citenamefont {Chen}, \citenamefont {Che}, \citenamefont
  {Yu}, \citenamefont {Zhang}, \citenamefont {Shao}, \citenamefont {Stern},
  \citenamefont {Casas}, \citenamefont {Xia}, \citenamefont {Han},
  \citenamefont {Kirby}, \citenamefont {Lake}, \citenamefont {Law},\ and\
  \citenamefont {Wang}}]{PhysRevLett.121.096802}%
  \BibitemOpen
  \bibfield  {author} {\bibinfo {author} {\bibfnamefont {Q.~L.}\ \bibnamefont
  {He}}, \bibinfo {author} {\bibfnamefont {G.}~\bibnamefont {Yin}}, \bibinfo
  {author} {\bibfnamefont {L.}~\bibnamefont {Yu}}, \bibinfo {author}
  {\bibfnamefont {A.~J.}\ \bibnamefont {Grutter}}, \bibinfo {author}
  {\bibfnamefont {L.}~\bibnamefont {Pan}}, \bibinfo {author} {\bibfnamefont
  {C.-Z.}\ \bibnamefont {Chen}}, \bibinfo {author} {\bibfnamefont
  {X.}~\bibnamefont {Che}}, \bibinfo {author} {\bibfnamefont {G.}~\bibnamefont
  {Yu}}, \bibinfo {author} {\bibfnamefont {B.}~\bibnamefont {Zhang}}, \bibinfo
  {author} {\bibfnamefont {Q.}~\bibnamefont {Shao}}, \bibinfo {author}
  {\bibfnamefont {A.~L.}\ \bibnamefont {Stern}}, \bibinfo {author}
  {\bibfnamefont {B.}~\bibnamefont {Casas}}, \bibinfo {author} {\bibfnamefont
  {J.}~\bibnamefont {Xia}}, \bibinfo {author} {\bibfnamefont {X.}~\bibnamefont
  {Han}}, \bibinfo {author} {\bibfnamefont {B.~J.}\ \bibnamefont {Kirby}},
  \bibinfo {author} {\bibfnamefont {R.~K.}\ \bibnamefont {Lake}}, \bibinfo
  {author} {\bibfnamefont {K.~T.}\ \bibnamefont {Law}}, \ and\ \bibinfo
  {author} {\bibfnamefont {K.~L.}\ \bibnamefont {Wang}},\ }\href {\doibase
  10.1103/PhysRevLett.121.096802} {\bibfield  {journal} {\bibinfo  {journal}
  {Phys. Rev. Lett.}\ }\textbf {\bibinfo {volume} {121}},\ \bibinfo {pages}
  {096802} (\bibinfo {year} {2018})}\BibitemShut {NoStop}%
\bibitem [{\citenamefont {Liu}\ \emph {et~al.}(2017)\citenamefont {Liu},
  \citenamefont {Zang}, \citenamefont {Ruan}, \citenamefont {Gong},
  \citenamefont {He}, \citenamefont {Ma}, \citenamefont {Xue},\ and\
  \citenamefont {Wang}}]{PhysRevLett.119.176809}%
  \BibitemOpen
  \bibfield  {author} {\bibinfo {author} {\bibfnamefont {C.}~\bibnamefont
  {Liu}}, \bibinfo {author} {\bibfnamefont {Y.}~\bibnamefont {Zang}}, \bibinfo
  {author} {\bibfnamefont {W.}~\bibnamefont {Ruan}}, \bibinfo {author}
  {\bibfnamefont {Y.}~\bibnamefont {Gong}}, \bibinfo {author} {\bibfnamefont
  {K.}~\bibnamefont {He}}, \bibinfo {author} {\bibfnamefont {X.}~\bibnamefont
  {Ma}}, \bibinfo {author} {\bibfnamefont {Q.-K.}\ \bibnamefont {Xue}}, \ and\
  \bibinfo {author} {\bibfnamefont {Y.}~\bibnamefont {Wang}},\ }\href {\doibase
  10.1103/PhysRevLett.119.176809} {\bibfield  {journal} {\bibinfo  {journal}
  {Phys. Rev. Lett.}\ }\textbf {\bibinfo {volume} {119}},\ \bibinfo {pages}
  {176809} (\bibinfo {year} {2017})}\BibitemShut {NoStop}%
\bibitem [{\citenamefont {Zhang}\ \emph {et~al.}(2018)\citenamefont {Zhang},
  \citenamefont {Kronast}, \citenamefont {van~der Laan},\ and\ \citenamefont
  {Hesjedal}}]{Zhang2018}%
  \BibitemOpen
  \bibfield  {author} {\bibinfo {author} {\bibfnamefont {S.}~\bibnamefont
  {Zhang}}, \bibinfo {author} {\bibfnamefont {F.}~\bibnamefont {Kronast}},
  \bibinfo {author} {\bibfnamefont {G.}~\bibnamefont {van~der Laan}}, \ and\
  \bibinfo {author} {\bibfnamefont {T.}~\bibnamefont {Hesjedal}},\ }\href
  {\doibase 10.1021/acs.nanolett.7b04537} {\bibfield  {journal} {\bibinfo
  {journal} {Nano Letters}\ }\textbf {\bibinfo {volume} {18}},\ \bibinfo
  {pages} {1057} (\bibinfo {year} {2018})}\BibitemShut {NoStop}%
\bibitem [{\citenamefont {Chen}\ \emph {et~al.}(2019)\citenamefont {Chen},
  \citenamefont {Wang}, \citenamefont {Zhang}, \citenamefont {Zhou},
  \citenamefont {Zhang}, \citenamefont {Jin}, \citenamefont {Wang},
  \citenamefont {Qin}, \citenamefont {Qiu}, \citenamefont {Mei}, \citenamefont
  {Ye}, \citenamefont {Xi}, \citenamefont {He}, \citenamefont {Li},\ and\
  \citenamefont {Wang}}]{Chen2019}%
  \BibitemOpen
  \bibfield  {author} {\bibinfo {author} {\bibfnamefont {J.}~\bibnamefont
  {Chen}}, \bibinfo {author} {\bibfnamefont {L.}~\bibnamefont {Wang}}, \bibinfo
  {author} {\bibfnamefont {M.}~\bibnamefont {Zhang}}, \bibinfo {author}
  {\bibfnamefont {L.}~\bibnamefont {Zhou}}, \bibinfo {author} {\bibfnamefont
  {R.}~\bibnamefont {Zhang}}, \bibinfo {author} {\bibfnamefont
  {L.}~\bibnamefont {Jin}}, \bibinfo {author} {\bibfnamefont {X.}~\bibnamefont
  {Wang}}, \bibinfo {author} {\bibfnamefont {H.}~\bibnamefont {Qin}}, \bibinfo
  {author} {\bibfnamefont {Y.}~\bibnamefont {Qiu}}, \bibinfo {author}
  {\bibfnamefont {J.}~\bibnamefont {Mei}}, \bibinfo {author} {\bibfnamefont
  {F.}~\bibnamefont {Ye}}, \bibinfo {author} {\bibfnamefont {B.}~\bibnamefont
  {Xi}}, \bibinfo {author} {\bibfnamefont {H.}~\bibnamefont {He}}, \bibinfo
  {author} {\bibfnamefont {B.}~\bibnamefont {Li}}, \ and\ \bibinfo {author}
  {\bibfnamefont {G.}~\bibnamefont {Wang}},\ }\href@noop {} {\bibfield
  {journal} {\bibinfo  {journal} {Nano Letters}\ }\textbf {\bibinfo {volume}
  {19}},\ \bibinfo {pages} {6144} (\bibinfo {year} {2019})}\BibitemShut
  {NoStop}%
\bibitem [{\citenamefont {Li}\ \emph {et~al.}(2021)\citenamefont {Li},
  \citenamefont {Ding}, \citenamefont {Zhang}, \citenamefont {Kally},
  \citenamefont {Pillsbury}, \citenamefont {Heinonen}, \citenamefont {Rimal},
  \citenamefont {Bi}, \citenamefont {DeMann}, \citenamefont {Field},
  \citenamefont {Wang}, \citenamefont {Tang}, \citenamefont {Jiang},
  \citenamefont {Hoffmann}, \citenamefont {Samarth},\ and\ \citenamefont
  {Wu}}]{Li2021}%
  \BibitemOpen
  \bibfield  {author} {\bibinfo {author} {\bibfnamefont {P.}~\bibnamefont
  {Li}}, \bibinfo {author} {\bibfnamefont {J.}~\bibnamefont {Ding}}, \bibinfo
  {author} {\bibfnamefont {S.~S.-L.}\ \bibnamefont {Zhang}}, \bibinfo {author}
  {\bibfnamefont {J.}~\bibnamefont {Kally}}, \bibinfo {author} {\bibfnamefont
  {T.}~\bibnamefont {Pillsbury}}, \bibinfo {author} {\bibfnamefont {O.~G.}\
  \bibnamefont {Heinonen}}, \bibinfo {author} {\bibfnamefont {G.}~\bibnamefont
  {Rimal}}, \bibinfo {author} {\bibfnamefont {C.}~\bibnamefont {Bi}}, \bibinfo
  {author} {\bibfnamefont {A.}~\bibnamefont {DeMann}}, \bibinfo {author}
  {\bibfnamefont {S.~B.}\ \bibnamefont {Field}}, \bibinfo {author}
  {\bibfnamefont {W.}~\bibnamefont {Wang}}, \bibinfo {author} {\bibfnamefont
  {J.}~\bibnamefont {Tang}}, \bibinfo {author} {\bibfnamefont {J.~S.}\
  \bibnamefont {Jiang}}, \bibinfo {author} {\bibfnamefont {A.}~\bibnamefont
  {Hoffmann}}, \bibinfo {author} {\bibfnamefont {N.}~\bibnamefont {Samarth}}, \
  and\ \bibinfo {author} {\bibfnamefont {M.}~\bibnamefont {Wu}},\ }\href@noop
  {} {\bibfield  {journal} {\bibinfo  {journal} {Nano Letters}\ }\textbf
  {\bibinfo {volume} {21}},\ \bibinfo {pages} {84} (\bibinfo {year}
  {2021})}\BibitemShut {NoStop}%
\bibitem [{\citenamefont {Nogueira}\ \emph {et~al.}(2018)\citenamefont
  {Nogueira}, \citenamefont {Eremin}, \citenamefont {Katmis}, \citenamefont
  {Moodera}, \citenamefont {van~den Brink},\ and\ \citenamefont
  {Kravchuk}}]{PhysRevB.98.060401}%
  \BibitemOpen
  \bibfield  {author} {\bibinfo {author} {\bibfnamefont {F.~S.}\ \bibnamefont
  {Nogueira}}, \bibinfo {author} {\bibfnamefont {I.}~\bibnamefont {Eremin}},
  \bibinfo {author} {\bibfnamefont {F.}~\bibnamefont {Katmis}}, \bibinfo
  {author} {\bibfnamefont {J.~S.}\ \bibnamefont {Moodera}}, \bibinfo {author}
  {\bibfnamefont {J.}~\bibnamefont {van~den Brink}}, \ and\ \bibinfo {author}
  {\bibfnamefont {V.~P.}\ \bibnamefont {Kravchuk}},\ }\href@noop {} {\bibfield
  {journal} {\bibinfo  {journal} {Phys. Rev. B}\ }\textbf {\bibinfo {volume}
  {98}},\ \bibinfo {pages} {060401} (\bibinfo {year} {2018})}\BibitemShut
  {NoStop}%
\bibitem [{\citenamefont {Che}\ \emph {et~al.}(2018)\citenamefont {Che},
  \citenamefont {Murata}, \citenamefont {Pan}, \citenamefont {He},
  \citenamefont {Yu}, \citenamefont {Shao}, \citenamefont {Yin}, \citenamefont
  {Deng}, \citenamefont {Fan}, \citenamefont {Ma}, \citenamefont {Liang},
  \citenamefont {Zhang}, \citenamefont {Han}, \citenamefont {Bi}, \citenamefont
  {Yang}, \citenamefont {Zhang},\ and\ \citenamefont {Wang}}]{Che2018}%
  \BibitemOpen
  \bibfield  {author} {\bibinfo {author} {\bibfnamefont {X.}~\bibnamefont
  {Che}}, \bibinfo {author} {\bibfnamefont {K.}~\bibnamefont {Murata}},
  \bibinfo {author} {\bibfnamefont {L.}~\bibnamefont {Pan}}, \bibinfo {author}
  {\bibfnamefont {Q.~L.}\ \bibnamefont {He}}, \bibinfo {author} {\bibfnamefont
  {G.}~\bibnamefont {Yu}}, \bibinfo {author} {\bibfnamefont {Q.}~\bibnamefont
  {Shao}}, \bibinfo {author} {\bibfnamefont {G.}~\bibnamefont {Yin}}, \bibinfo
  {author} {\bibfnamefont {P.}~\bibnamefont {Deng}}, \bibinfo {author}
  {\bibfnamefont {Y.}~\bibnamefont {Fan}}, \bibinfo {author} {\bibfnamefont
  {B.}~\bibnamefont {Ma}}, \bibinfo {author} {\bibfnamefont {X.}~\bibnamefont
  {Liang}}, \bibinfo {author} {\bibfnamefont {B.}~\bibnamefont {Zhang}},
  \bibinfo {author} {\bibfnamefont {X.}~\bibnamefont {Han}}, \bibinfo {author}
  {\bibfnamefont {L.}~\bibnamefont {Bi}}, \bibinfo {author} {\bibfnamefont
  {Q.-H.}\ \bibnamefont {Yang}}, \bibinfo {author} {\bibfnamefont
  {H.}~\bibnamefont {Zhang}}, \ and\ \bibinfo {author} {\bibfnamefont {K.~L.}\
  \bibnamefont {Wang}},\ }\href {\doibase 10.1021/acsnano.8b02647} {\bibfield
  {journal} {\bibinfo  {journal} {ACS Nano}\ }\textbf {\bibinfo {volume}
  {12}},\ \bibinfo {pages} {5042} (\bibinfo {year} {2018})}\BibitemShut
  {NoStop}%
\bibitem [{\citenamefont {Lang}\ \emph {et~al.}(2014)\citenamefont {Lang},
  \citenamefont {Montazeri}, \citenamefont {Onbasli}, \citenamefont {Kou},
  \citenamefont {Fan}, \citenamefont {Upadhyaya}, \citenamefont {Yao},
  \citenamefont {Liu}, \citenamefont {Jiang}, \citenamefont {Jiang},
  \citenamefont {Wong}, \citenamefont {Yu}, \citenamefont {Tang}, \citenamefont
  {Nie}, \citenamefont {He}, \citenamefont {Schwartz}, \citenamefont {Wang},
  \citenamefont {Ross},\ and\ \citenamefont {Wang}}]{Lang2014}%
  \BibitemOpen
  \bibfield  {author} {\bibinfo {author} {\bibfnamefont {M.}~\bibnamefont
  {Lang}}, \bibinfo {author} {\bibfnamefont {M.}~\bibnamefont {Montazeri}},
  \bibinfo {author} {\bibfnamefont {M.~C.}\ \bibnamefont {Onbasli}}, \bibinfo
  {author} {\bibfnamefont {X.}~\bibnamefont {Kou}}, \bibinfo {author}
  {\bibfnamefont {Y.}~\bibnamefont {Fan}}, \bibinfo {author} {\bibfnamefont
  {P.}~\bibnamefont {Upadhyaya}}, \bibinfo {author} {\bibfnamefont
  {K.}~\bibnamefont {Yao}}, \bibinfo {author} {\bibfnamefont {F.}~\bibnamefont
  {Liu}}, \bibinfo {author} {\bibfnamefont {Y.}~\bibnamefont {Jiang}}, \bibinfo
  {author} {\bibfnamefont {W.}~\bibnamefont {Jiang}}, \bibinfo {author}
  {\bibfnamefont {K.~L.}\ \bibnamefont {Wong}}, \bibinfo {author}
  {\bibfnamefont {G.}~\bibnamefont {Yu}}, \bibinfo {author} {\bibfnamefont
  {J.}~\bibnamefont {Tang}}, \bibinfo {author} {\bibfnamefont {T.}~\bibnamefont
  {Nie}}, \bibinfo {author} {\bibfnamefont {L.}~\bibnamefont {He}}, \bibinfo
  {author} {\bibfnamefont {R.~N.}\ \bibnamefont {Schwartz}}, \bibinfo {author}
  {\bibfnamefont {Y.}~\bibnamefont {Wang}}, \bibinfo {author} {\bibfnamefont
  {C.~A.}\ \bibnamefont {Ross}}, \ and\ \bibinfo {author} {\bibfnamefont
  {K.~L.}\ \bibnamefont {Wang}},\ }\href {\doibase 10.1021/nl500973k}
  {\bibfield  {journal} {\bibinfo  {journal} {Nano Letters}\ }\textbf {\bibinfo
  {volume} {14}},\ \bibinfo {pages} {3459} (\bibinfo {year}
  {2014})}\BibitemShut {NoStop}%
\bibitem [{\citenamefont {Tang}\ \emph {et~al.}(2017)\citenamefont {Tang},
  \citenamefont {Chang}, \citenamefont {Zhao}, \citenamefont {Liu},
  \citenamefont {Jiang}, \citenamefont {Liu}, \citenamefont {McCartney},
  \citenamefont {Smith}, \citenamefont {Chen}, \citenamefont {Moodera},\ and\
  \citenamefont {Shi}}]{Tange1700307}%
  \BibitemOpen
  \bibfield  {author} {\bibinfo {author} {\bibfnamefont {C.}~\bibnamefont
  {Tang}}, \bibinfo {author} {\bibfnamefont {C.-Z.}\ \bibnamefont {Chang}},
  \bibinfo {author} {\bibfnamefont {G.}~\bibnamefont {Zhao}}, \bibinfo {author}
  {\bibfnamefont {Y.}~\bibnamefont {Liu}}, \bibinfo {author} {\bibfnamefont
  {Z.}~\bibnamefont {Jiang}}, \bibinfo {author} {\bibfnamefont {C.-X.}\
  \bibnamefont {Liu}}, \bibinfo {author} {\bibfnamefont {M.~R.}\ \bibnamefont
  {McCartney}}, \bibinfo {author} {\bibfnamefont {D.~J.}\ \bibnamefont
  {Smith}}, \bibinfo {author} {\bibfnamefont {T.}~\bibnamefont {Chen}},
  \bibinfo {author} {\bibfnamefont {J.~S.}\ \bibnamefont {Moodera}}, \ and\
  \bibinfo {author} {\bibfnamefont {J.}~\bibnamefont {Shi}},\ }\href@noop {}
  {\bibfield  {journal} {\bibinfo  {journal} {Science Advances}\ }\textbf
  {\bibinfo {volume} {3}} (\bibinfo {year} {2017})}\BibitemShut {NoStop}%
\bibitem [{\citenamefont {Fanchiang}\ \emph {et~al.}(2018)\citenamefont
  {Fanchiang}, \citenamefont {Chen}, \citenamefont {Tseng}, \citenamefont
  {Chen}, \citenamefont {Cheng}, \citenamefont {Yang}, \citenamefont {Wu},
  \citenamefont {Lee}, \citenamefont {Hong},\ and\ \citenamefont
  {Kwo}}]{Fanchiang2018}%
  \BibitemOpen
  \bibfield  {author} {\bibinfo {author} {\bibfnamefont {Y.~T.}\ \bibnamefont
  {Fanchiang}}, \bibinfo {author} {\bibfnamefont {K.~H.~M.}\ \bibnamefont
  {Chen}}, \bibinfo {author} {\bibfnamefont {C.~C.}\ \bibnamefont {Tseng}},
  \bibinfo {author} {\bibfnamefont {C.~C.}\ \bibnamefont {Chen}}, \bibinfo
  {author} {\bibfnamefont {C.~K.}\ \bibnamefont {Cheng}}, \bibinfo {author}
  {\bibfnamefont {S.~R.}\ \bibnamefont {Yang}}, \bibinfo {author}
  {\bibfnamefont {C.~N.}\ \bibnamefont {Wu}}, \bibinfo {author} {\bibfnamefont
  {S.~F.}\ \bibnamefont {Lee}}, \bibinfo {author} {\bibfnamefont
  {M.}~\bibnamefont {Hong}}, \ and\ \bibinfo {author} {\bibfnamefont
  {J.}~\bibnamefont {Kwo}},\ }\href@noop {} {\bibfield  {journal} {\bibinfo
  {journal} {Nature Communications}\ }\textbf {\bibinfo {volume} {9}},\
  \bibinfo {pages} {223} (\bibinfo {year} {2018})}\BibitemShut {NoStop}%
\bibitem [{\citenamefont {Liang}\ \emph {et~al.}(2015)\citenamefont {Liang},
  \citenamefont {DeGrave}, \citenamefont {Stolt}, \citenamefont {Tokura},\ and\
  \citenamefont {Jin}}]{Liang2015}%
  \BibitemOpen
  \bibfield  {author} {\bibinfo {author} {\bibfnamefont {D.}~\bibnamefont
  {Liang}}, \bibinfo {author} {\bibfnamefont {J.~P.}\ \bibnamefont {DeGrave}},
  \bibinfo {author} {\bibfnamefont {M.~J.}\ \bibnamefont {Stolt}}, \bibinfo
  {author} {\bibfnamefont {Y.}~\bibnamefont {Tokura}}, \ and\ \bibinfo {author}
  {\bibfnamefont {S.}~\bibnamefont {Jin}},\ }\href@noop {} {\bibfield
  {journal} {\bibinfo  {journal} {Nature Communications}\ }\textbf {\bibinfo
  {volume} {6}},\ \bibinfo {pages} {8217} (\bibinfo {year} {2015})}\BibitemShut
  {NoStop}%
\bibitem [{\citenamefont {Katmis}\ \emph {et~al.}(2016)\citenamefont {Katmis},
  \citenamefont {Lauter}, \citenamefont {Nogueira}, \citenamefont {Assaf},
  \citenamefont {Jamer}, \citenamefont {Wei}, \citenamefont {Satpati},
  \citenamefont {Freeland}, \citenamefont {Eremin}, \citenamefont {Heiman},
  \citenamefont {Jarillo-Herrero},\ and\ \citenamefont {Moodera}}]{Katmis2016}%
  \BibitemOpen
  \bibfield  {author} {\bibinfo {author} {\bibfnamefont {F.}~\bibnamefont
  {Katmis}}, \bibinfo {author} {\bibfnamefont {V.}~\bibnamefont {Lauter}},
  \bibinfo {author} {\bibfnamefont {F.~S.}\ \bibnamefont {Nogueira}}, \bibinfo
  {author} {\bibfnamefont {B.~A.}\ \bibnamefont {Assaf}}, \bibinfo {author}
  {\bibfnamefont {M.~E.}\ \bibnamefont {Jamer}}, \bibinfo {author}
  {\bibfnamefont {P.}~\bibnamefont {Wei}}, \bibinfo {author} {\bibfnamefont
  {B.}~\bibnamefont {Satpati}}, \bibinfo {author} {\bibfnamefont {J.~W.}\
  \bibnamefont {Freeland}}, \bibinfo {author} {\bibfnamefont {I.}~\bibnamefont
  {Eremin}}, \bibinfo {author} {\bibfnamefont {D.}~\bibnamefont {Heiman}},
  \bibinfo {author} {\bibfnamefont {P.}~\bibnamefont {Jarillo-Herrero}}, \ and\
  \bibinfo {author} {\bibfnamefont {J.~S.}\ \bibnamefont {Moodera}},\ }\href
  {\doibase 10.1038/nature17635} {\bibfield  {journal} {\bibinfo  {journal}
  {Nature}\ }\textbf {\bibinfo {volume} {533}},\ \bibinfo {pages} {513}
  (\bibinfo {year} {2016})}\BibitemShut {NoStop}%
\bibitem [{\citenamefont {Wei}\ \emph {et~al.}(2013)\citenamefont {Wei},
  \citenamefont {Katmis}, \citenamefont {Assaf}, \citenamefont {Steinberg},
  \citenamefont {Jarillo-Herrero}, \citenamefont {Heiman},\ and\ \citenamefont
  {Moodera}}]{PhysRevLett.110.186807}%
  \BibitemOpen
  \bibfield  {author} {\bibinfo {author} {\bibfnamefont {P.}~\bibnamefont
  {Wei}}, \bibinfo {author} {\bibfnamefont {F.}~\bibnamefont {Katmis}},
  \bibinfo {author} {\bibfnamefont {B.~A.}\ \bibnamefont {Assaf}}, \bibinfo
  {author} {\bibfnamefont {H.}~\bibnamefont {Steinberg}}, \bibinfo {author}
  {\bibfnamefont {P.}~\bibnamefont {Jarillo-Herrero}}, \bibinfo {author}
  {\bibfnamefont {D.}~\bibnamefont {Heiman}}, \ and\ \bibinfo {author}
  {\bibfnamefont {J.~S.}\ \bibnamefont {Moodera}},\ }\href {\doibase
  10.1103/PhysRevLett.110.186807} {\bibfield  {journal} {\bibinfo  {journal}
  {Phys. Rev. Lett.}\ }\textbf {\bibinfo {volume} {110}},\ \bibinfo {pages}
  {186807} (\bibinfo {year} {2013})}\BibitemShut {NoStop}%
\bibitem [{\citenamefont {Kim}\ \emph {et~al.}(2017)\citenamefont {Kim},
  \citenamefont {Kim}, \citenamefont {Wang}, \citenamefont {Sinova},\ and\
  \citenamefont {Wu}}]{PhysRevLett.119.027201}%
  \BibitemOpen
  \bibfield  {author} {\bibinfo {author} {\bibfnamefont {J.}~\bibnamefont
  {Kim}}, \bibinfo {author} {\bibfnamefont {K.-W.}\ \bibnamefont {Kim}},
  \bibinfo {author} {\bibfnamefont {H.}~\bibnamefont {Wang}}, \bibinfo {author}
  {\bibfnamefont {J.}~\bibnamefont {Sinova}}, \ and\ \bibinfo {author}
  {\bibfnamefont {R.}~\bibnamefont {Wu}},\ }\href {\doibase
  10.1103/PhysRevLett.119.027201} {\bibfield  {journal} {\bibinfo  {journal}
  {Phys. Rev. Lett.}\ }\textbf {\bibinfo {volume} {119}},\ \bibinfo {pages}
  {027201} (\bibinfo {year} {2017})}\BibitemShut {NoStop}%
\bibitem [{\citenamefont {Krieger}\ \emph {et~al.}(2019)\citenamefont
  {Krieger}, \citenamefont {Ou}, \citenamefont {Caputo}, \citenamefont
  {Chikina}, \citenamefont {D\"obeli}, \citenamefont {Husanu}, \citenamefont
  {Keren}, \citenamefont {Prokscha}, \citenamefont {Suter}, \citenamefont
  {Chang}, \citenamefont {Moodera}, \citenamefont {Strocov},\ and\
  \citenamefont {Salman}}]{PhysRevB.99.064423}%
  \BibitemOpen
  \bibfield  {author} {\bibinfo {author} {\bibfnamefont {J.~A.}\ \bibnamefont
  {Krieger}}, \bibinfo {author} {\bibfnamefont {Y.}~\bibnamefont {Ou}},
  \bibinfo {author} {\bibfnamefont {M.}~\bibnamefont {Caputo}}, \bibinfo
  {author} {\bibfnamefont {A.}~\bibnamefont {Chikina}}, \bibinfo {author}
  {\bibfnamefont {M.}~\bibnamefont {D\"obeli}}, \bibinfo {author}
  {\bibfnamefont {M.-A.}\ \bibnamefont {Husanu}}, \bibinfo {author}
  {\bibfnamefont {I.}~\bibnamefont {Keren}}, \bibinfo {author} {\bibfnamefont
  {T.}~\bibnamefont {Prokscha}}, \bibinfo {author} {\bibfnamefont
  {A.}~\bibnamefont {Suter}}, \bibinfo {author} {\bibfnamefont {C.-Z.}\
  \bibnamefont {Chang}}, \bibinfo {author} {\bibfnamefont {J.~S.}\ \bibnamefont
  {Moodera}}, \bibinfo {author} {\bibfnamefont {V.~N.}\ \bibnamefont
  {Strocov}}, \ and\ \bibinfo {author} {\bibfnamefont {Z.}~\bibnamefont
  {Salman}},\ }\href {\doibase 10.1103/PhysRevB.99.064423} {\bibfield
  {journal} {\bibinfo  {journal} {Phys. Rev. B}\ }\textbf {\bibinfo {volume}
  {99}},\ \bibinfo {pages} {064423} (\bibinfo {year} {2019})}\BibitemShut
  {NoStop}%
\bibitem [{\citenamefont {Wei}\ \emph {et~al.}(2016)\citenamefont {Wei},
  \citenamefont {Lee}, \citenamefont {Lemaitre}, \citenamefont {Pinel},
  \citenamefont {Cutaia}, \citenamefont {Cha}, \citenamefont {Katmis},
  \citenamefont {Zhu}, \citenamefont {Heiman}, \citenamefont {Hone},
  \citenamefont {Moodera},\ and\ \citenamefont {Chen}}]{Wei2016}%
  \BibitemOpen
  \bibfield  {author} {\bibinfo {author} {\bibfnamefont {P.}~\bibnamefont
  {Wei}}, \bibinfo {author} {\bibfnamefont {S.}~\bibnamefont {Lee}}, \bibinfo
  {author} {\bibfnamefont {F.}~\bibnamefont {Lemaitre}}, \bibinfo {author}
  {\bibfnamefont {L.}~\bibnamefont {Pinel}}, \bibinfo {author} {\bibfnamefont
  {D.}~\bibnamefont {Cutaia}}, \bibinfo {author} {\bibfnamefont
  {W.}~\bibnamefont {Cha}}, \bibinfo {author} {\bibfnamefont {F.}~\bibnamefont
  {Katmis}}, \bibinfo {author} {\bibfnamefont {Y.}~\bibnamefont {Zhu}},
  \bibinfo {author} {\bibfnamefont {D.}~\bibnamefont {Heiman}}, \bibinfo
  {author} {\bibfnamefont {J.}~\bibnamefont {Hone}}, \bibinfo {author}
  {\bibfnamefont {J.~S.}\ \bibnamefont {Moodera}}, \ and\ \bibinfo {author}
  {\bibfnamefont {C.-T.}\ \bibnamefont {Chen}},\ }\href {\doibase
  10.1038/nmat4603} {\bibfield  {journal} {\bibinfo  {journal} {Nature
  Materials}\ }\textbf {\bibinfo {volume} {15}},\ \bibinfo {pages} {711}
  (\bibinfo {year} {2016})}\BibitemShut {NoStop}%
\bibitem [{\citenamefont {Mathimalar}\ \emph {et~al.}(2020)\citenamefont
  {Mathimalar}, \citenamefont {Sasmal}, \citenamefont {Bhardwaj}, \citenamefont
  {Abhaya}, \citenamefont {Pothala}, \citenamefont {Chaudhary}, \citenamefont
  {Satpati},\ and\ \citenamefont {Raman}}]{Mathimalar2020}%
  \BibitemOpen
  \bibfield  {author} {\bibinfo {author} {\bibfnamefont {S.}~\bibnamefont
  {Mathimalar}}, \bibinfo {author} {\bibfnamefont {S.}~\bibnamefont {Sasmal}},
  \bibinfo {author} {\bibfnamefont {A.}~\bibnamefont {Bhardwaj}}, \bibinfo
  {author} {\bibfnamefont {S.}~\bibnamefont {Abhaya}}, \bibinfo {author}
  {\bibfnamefont {R.}~\bibnamefont {Pothala}}, \bibinfo {author} {\bibfnamefont
  {S.}~\bibnamefont {Chaudhary}}, \bibinfo {author} {\bibfnamefont
  {B.}~\bibnamefont {Satpati}}, \ and\ \bibinfo {author} {\bibfnamefont
  {K.~V.}\ \bibnamefont {Raman}},\ }\href {\doibase 10.1038/s41535-020-00267-5}
  {\bibfield  {journal} {\bibinfo  {journal} {npj Quantum Materials}\ }\textbf
  {\bibinfo {volume} {5}},\ \bibinfo {pages} {64} (\bibinfo {year}
  {2020})}\BibitemShut {NoStop}%
\bibitem [{\citenamefont {Moon}\ \emph {et~al.}(2019)\citenamefont {Moon},
  \citenamefont {Yoon}, \citenamefont {Kim},\ and\ \citenamefont
  {Hwang}}]{PhysRevApplied.12.064054}%
  \BibitemOpen
  \bibfield  {author} {\bibinfo {author} {\bibfnamefont {K.-W.}\ \bibnamefont
  {Moon}}, \bibinfo {author} {\bibfnamefont {J.}~\bibnamefont {Yoon}}, \bibinfo
  {author} {\bibfnamefont {C.}~\bibnamefont {Kim}}, \ and\ \bibinfo {author}
  {\bibfnamefont {C.}~\bibnamefont {Hwang}},\ }\href {\doibase
  10.1103/PhysRevApplied.12.064054} {\bibfield  {journal} {\bibinfo  {journal}
  {Phys. Rev. Applied}\ }\textbf {\bibinfo {volume} {12}},\ \bibinfo {pages}
  {064054} (\bibinfo {year} {2019})}\BibitemShut {NoStop}%
\bibitem [{\citenamefont {Li}\ \emph {et~al.}(2017)\citenamefont {Li},
  \citenamefont {Song}, \citenamefont {Zhao}, \citenamefont {Garlow},
  \citenamefont {Liu}, \citenamefont {Wu}, \citenamefont {Zhu}, \citenamefont
  {Moodera}, \citenamefont {Chan}, \citenamefont {Chen},\ and\ \citenamefont
  {Chang}}]{PhysRevB.96.201301}%
  \BibitemOpen
  \bibfield  {author} {\bibinfo {author} {\bibfnamefont {M.}~\bibnamefont
  {Li}}, \bibinfo {author} {\bibfnamefont {Q.}~\bibnamefont {Song}}, \bibinfo
  {author} {\bibfnamefont {W.}~\bibnamefont {Zhao}}, \bibinfo {author}
  {\bibfnamefont {J.~A.}\ \bibnamefont {Garlow}}, \bibinfo {author}
  {\bibfnamefont {T.-H.}\ \bibnamefont {Liu}}, \bibinfo {author} {\bibfnamefont
  {L.}~\bibnamefont {Wu}}, \bibinfo {author} {\bibfnamefont {Y.}~\bibnamefont
  {Zhu}}, \bibinfo {author} {\bibfnamefont {J.~S.}\ \bibnamefont {Moodera}},
  \bibinfo {author} {\bibfnamefont {M.~H.~W.}\ \bibnamefont {Chan}}, \bibinfo
  {author} {\bibfnamefont {G.}~\bibnamefont {Chen}}, \ and\ \bibinfo {author}
  {\bibfnamefont {C.-Z.}\ \bibnamefont {Chang}},\ }\href {\doibase
  10.1103/PhysRevB.96.201301} {\bibfield  {journal} {\bibinfo  {journal} {Phys.
  Rev. B}\ }\textbf {\bibinfo {volume} {96}},\ \bibinfo {pages} {201301}
  (\bibinfo {year} {2017})}\BibitemShut {NoStop}%
\bibitem [{\citenamefont {Du}\ \emph {et~al.}(2014)\citenamefont {Du},
  \citenamefont {DeGrave}, \citenamefont {Xue}, \citenamefont {Liang},
  \citenamefont {Ning}, \citenamefont {Yang}, \citenamefont {Tian},
  \citenamefont {Zhang},\ and\ \citenamefont {Jin}}]{Du2014}%
  \BibitemOpen
  \bibfield  {author} {\bibinfo {author} {\bibfnamefont {H.}~\bibnamefont
  {Du}}, \bibinfo {author} {\bibfnamefont {J.~P.}\ \bibnamefont {DeGrave}},
  \bibinfo {author} {\bibfnamefont {F.}~\bibnamefont {Xue}}, \bibinfo {author}
  {\bibfnamefont {D.}~\bibnamefont {Liang}}, \bibinfo {author} {\bibfnamefont
  {W.}~\bibnamefont {Ning}}, \bibinfo {author} {\bibfnamefont {J.}~\bibnamefont
  {Yang}}, \bibinfo {author} {\bibfnamefont {M.}~\bibnamefont {Tian}}, \bibinfo
  {author} {\bibfnamefont {Y.}~\bibnamefont {Zhang}}, \ and\ \bibinfo {author}
  {\bibfnamefont {S.}~\bibnamefont {Jin}},\ }\href {\doibase 10.1021/nl5001899}
  {\bibfield  {journal} {\bibinfo  {journal} {Nano Letters}\ }\textbf {\bibinfo
  {volume} {14}},\ \bibinfo {pages} {2026} (\bibinfo {year}
  {2014})}\BibitemShut {NoStop}%
\bibitem [{\citenamefont {Rakhmilevich}\ \emph {et~al.}(2018)\citenamefont
  {Rakhmilevich}, \citenamefont {Wang}, \citenamefont {Zhao}, \citenamefont
  {Chan}, \citenamefont {Moodera}, \citenamefont {Liu},\ and\ \citenamefont
  {Chang}}]{PhysRevB.98.094404}%
  \BibitemOpen
  \bibfield  {author} {\bibinfo {author} {\bibfnamefont {D.}~\bibnamefont
  {Rakhmilevich}}, \bibinfo {author} {\bibfnamefont {F.}~\bibnamefont {Wang}},
  \bibinfo {author} {\bibfnamefont {W.}~\bibnamefont {Zhao}}, \bibinfo {author}
  {\bibfnamefont {M.~H.~W.}\ \bibnamefont {Chan}}, \bibinfo {author}
  {\bibfnamefont {J.~S.}\ \bibnamefont {Moodera}}, \bibinfo {author}
  {\bibfnamefont {C.}~\bibnamefont {Liu}}, \ and\ \bibinfo {author}
  {\bibfnamefont {C.-Z.}\ \bibnamefont {Chang}},\ }\href {\doibase
  10.1103/PhysRevB.98.094404} {\bibfield  {journal} {\bibinfo  {journal} {Phys.
  Rev. B}\ }\textbf {\bibinfo {volume} {98}},\ \bibinfo {pages} {094404}
  (\bibinfo {year} {2018})}\BibitemShut {NoStop}%
\bibitem [{\citenamefont {Taskin}\ \emph {et~al.}(2017)\citenamefont {Taskin},
  \citenamefont {Legg}, \citenamefont {Yang}, \citenamefont {Sasaki},
  \citenamefont {Kanai}, \citenamefont {Matsumoto}, \citenamefont {Rosch},\
  and\ \citenamefont {Ando}}]{Taskin2017}%
  \BibitemOpen
  \bibfield  {author} {\bibinfo {author} {\bibfnamefont {A.~A.}\ \bibnamefont
  {Taskin}}, \bibinfo {author} {\bibfnamefont {H.~F.}\ \bibnamefont {Legg}},
  \bibinfo {author} {\bibfnamefont {F.}~\bibnamefont {Yang}}, \bibinfo {author}
  {\bibfnamefont {S.}~\bibnamefont {Sasaki}}, \bibinfo {author} {\bibfnamefont
  {Y.}~\bibnamefont {Kanai}}, \bibinfo {author} {\bibfnamefont
  {K.}~\bibnamefont {Matsumoto}}, \bibinfo {author} {\bibfnamefont
  {A.}~\bibnamefont {Rosch}}, \ and\ \bibinfo {author} {\bibfnamefont
  {Y.}~\bibnamefont {Ando}},\ }\href {\doibase 10.1038/s41467-017-01474-8}
  {\bibfield  {journal} {\bibinfo  {journal} {Nature Communications}\ }\textbf
  {\bibinfo {volume} {8}},\ \bibinfo {pages} {1340} (\bibinfo {year}
  {2017})}\BibitemShut {NoStop}%
\bibitem [{\citenamefont {Bhardwaj}\ \emph {et~al.}(2021)\citenamefont
  {Bhardwaj}, \citenamefont {Prasad~P.}, \citenamefont {Raman},\ and\
  \citenamefont {Suri}}]{bhardwaj2021observation}%
  \BibitemOpen
  \bibfield  {author} {\bibinfo {author} {\bibfnamefont {A.}~\bibnamefont
  {Bhardwaj}}, \bibinfo {author} {\bibfnamefont {S.}~\bibnamefont {Prasad~P.}},
  \bibinfo {author} {\bibfnamefont {K.~V.}\ \bibnamefont {Raman}}, \ and\
  \bibinfo {author} {\bibfnamefont {D.}~\bibnamefont {Suri}},\ }\href@noop {}
  {\bibfield  {journal} {\bibinfo  {journal} {Applied Physics Letters}\
  }\textbf {\bibinfo {volume} {118}},\ \bibinfo {pages} {241901} (\bibinfo
  {year} {2021})}\BibitemShut {NoStop}%
\bibitem [{\citenamefont {Yokouchi}\ \emph {et~al.}(2015)\citenamefont
  {Yokouchi}, \citenamefont {Kanazawa}, \citenamefont {Tsukazaki},
  \citenamefont {Kozuka}, \citenamefont {Kikkawa}, \citenamefont {Taguchi},
  \citenamefont {Kawasaki}, \citenamefont {Ichikawa}, \citenamefont {Kagawa},\
  and\ \citenamefont {Tokura}}]{doi:10.7566/JPSJ.84.104708}%
  \BibitemOpen
  \bibfield  {author} {\bibinfo {author} {\bibfnamefont {T.}~\bibnamefont
  {Yokouchi}}, \bibinfo {author} {\bibfnamefont {N.}~\bibnamefont {Kanazawa}},
  \bibinfo {author} {\bibfnamefont {A.}~\bibnamefont {Tsukazaki}}, \bibinfo
  {author} {\bibfnamefont {Y.}~\bibnamefont {Kozuka}}, \bibinfo {author}
  {\bibfnamefont {A.}~\bibnamefont {Kikkawa}}, \bibinfo {author} {\bibfnamefont
  {Y.}~\bibnamefont {Taguchi}}, \bibinfo {author} {\bibfnamefont
  {M.}~\bibnamefont {Kawasaki}}, \bibinfo {author} {\bibfnamefont
  {M.}~\bibnamefont {Ichikawa}}, \bibinfo {author} {\bibfnamefont
  {F.}~\bibnamefont {Kagawa}}, \ and\ \bibinfo {author} {\bibfnamefont
  {Y.}~\bibnamefont {Tokura}},\ }\href@noop {} {\bibfield  {journal} {\bibinfo
  {journal} {Journal of the Physical Society of Japan}\ }\textbf {\bibinfo
  {volume} {84}},\ \bibinfo {pages} {104708} (\bibinfo {year}
  {2015})}\BibitemShut {NoStop}%
\bibitem [{\citenamefont {Mathur}\ \emph {et~al.}(2021)\citenamefont {Mathur},
  \citenamefont {Yasin}, \citenamefont {Stolt}, \citenamefont {Nagai},
  \citenamefont {Kimoto}, \citenamefont {Du}, \citenamefont {Tian},
  \citenamefont {Tokura}, \citenamefont {Yu},\ and\ \citenamefont
  {Jin}}]{https://doi.org/10.1002/adfm.202008521}%
  \BibitemOpen
  \bibfield  {author} {\bibinfo {author} {\bibfnamefont {N.}~\bibnamefont
  {Mathur}}, \bibinfo {author} {\bibfnamefont {F.~S.}\ \bibnamefont {Yasin}},
  \bibinfo {author} {\bibfnamefont {M.~J.}\ \bibnamefont {Stolt}}, \bibinfo
  {author} {\bibfnamefont {T.}~\bibnamefont {Nagai}}, \bibinfo {author}
  {\bibfnamefont {K.}~\bibnamefont {Kimoto}}, \bibinfo {author} {\bibfnamefont
  {H.}~\bibnamefont {Du}}, \bibinfo {author} {\bibfnamefont {M.}~\bibnamefont
  {Tian}}, \bibinfo {author} {\bibfnamefont {Y.}~\bibnamefont {Tokura}},
  \bibinfo {author} {\bibfnamefont {X.}~\bibnamefont {Yu}}, \ and\ \bibinfo
  {author} {\bibfnamefont {S.}~\bibnamefont {Jin}},\ }\href@noop {} {\bibfield
  {journal} {\bibinfo  {journal} {Advanced Functional Materials}\ }\textbf
  {\bibinfo {volume} {31}},\ \bibinfo {pages} {2008521} (\bibinfo {year}
  {2021})}\BibitemShut {NoStop}%
\bibitem [{\citenamefont {Yokouchi}\ \emph {et~al.}(2018)\citenamefont
  {Yokouchi}, \citenamefont {Hoshino}, \citenamefont {Kanazawa}, \citenamefont
  {Kikkawa}, \citenamefont {Morikawa}, \citenamefont {Shibata}, \citenamefont
  {Arima}, \citenamefont {Taguchi}, \citenamefont {Kagawa}, \citenamefont
  {Nagaosa},\ and\ \citenamefont {Tokura}}]{Yokouchieaat1115}%
  \BibitemOpen
  \bibfield  {author} {\bibinfo {author} {\bibfnamefont {T.}~\bibnamefont
  {Yokouchi}}, \bibinfo {author} {\bibfnamefont {S.}~\bibnamefont {Hoshino}},
  \bibinfo {author} {\bibfnamefont {N.}~\bibnamefont {Kanazawa}}, \bibinfo
  {author} {\bibfnamefont {A.}~\bibnamefont {Kikkawa}}, \bibinfo {author}
  {\bibfnamefont {D.}~\bibnamefont {Morikawa}}, \bibinfo {author}
  {\bibfnamefont {K.}~\bibnamefont {Shibata}}, \bibinfo {author} {\bibfnamefont
  {T.-h.}\ \bibnamefont {Arima}}, \bibinfo {author} {\bibfnamefont
  {Y.}~\bibnamefont {Taguchi}}, \bibinfo {author} {\bibfnamefont
  {F.}~\bibnamefont {Kagawa}}, \bibinfo {author} {\bibfnamefont
  {N.}~\bibnamefont {Nagaosa}}, \ and\ \bibinfo {author} {\bibfnamefont
  {Y.}~\bibnamefont {Tokura}},\ }\href {\doibase 10.1126/sciadv.aat1115}
  {\bibfield  {journal} {\bibinfo  {journal} {Science Advances}\ }\textbf
  {\bibinfo {volume} {4}} (\bibinfo {year} {2018}),\
  10.1126/sciadv.aat1115}\BibitemShut {NoStop}%
\bibitem [{\citenamefont {Ritz}\ \emph {et~al.}(2013)\citenamefont {Ritz},
  \citenamefont {Halder}, \citenamefont {Franz}, \citenamefont {Bauer},
  \citenamefont {Wagner}, \citenamefont {Bamler}, \citenamefont {Rosch},\ and\
  \citenamefont {Pfleiderer}}]{PhysRevB.87.134424}%
  \BibitemOpen
  \bibfield  {author} {\bibinfo {author} {\bibfnamefont {R.}~\bibnamefont
  {Ritz}}, \bibinfo {author} {\bibfnamefont {M.}~\bibnamefont {Halder}},
  \bibinfo {author} {\bibfnamefont {C.}~\bibnamefont {Franz}}, \bibinfo
  {author} {\bibfnamefont {A.}~\bibnamefont {Bauer}}, \bibinfo {author}
  {\bibfnamefont {M.}~\bibnamefont {Wagner}}, \bibinfo {author} {\bibfnamefont
  {R.}~\bibnamefont {Bamler}}, \bibinfo {author} {\bibfnamefont
  {A.}~\bibnamefont {Rosch}}, \ and\ \bibinfo {author} {\bibfnamefont
  {C.}~\bibnamefont {Pfleiderer}},\ }\href {\doibase
  10.1103/PhysRevB.87.134424} {\bibfield  {journal} {\bibinfo  {journal} {Phys.
  Rev. B}\ }\textbf {\bibinfo {volume} {87}},\ \bibinfo {pages} {134424}
  (\bibinfo {year} {2013})}\BibitemShut {NoStop}%
\bibitem [{\citenamefont {Liu}\ \emph {et~al.}(2020)\citenamefont {Liu},
  \citenamefont {Zuo}, \citenamefont {Li}, \citenamefont {Liu}, \citenamefont
  {Zheng}, \citenamefont {Zhang}, \citenamefont {Zhao}, \citenamefont {Hu},
  \citenamefont {Sun},\ and\ \citenamefont {Shen}}]{LIU2020268}%
  \BibitemOpen
  \bibfield  {author} {\bibinfo {author} {\bibfnamefont {J.}~\bibnamefont
  {Liu}}, \bibinfo {author} {\bibfnamefont {S.}~\bibnamefont {Zuo}}, \bibinfo
  {author} {\bibfnamefont {H.}~\bibnamefont {Li}}, \bibinfo {author}
  {\bibfnamefont {Y.}~\bibnamefont {Liu}}, \bibinfo {author} {\bibfnamefont
  {X.}~\bibnamefont {Zheng}}, \bibinfo {author} {\bibfnamefont
  {Y.}~\bibnamefont {Zhang}}, \bibinfo {author} {\bibfnamefont
  {T.}~\bibnamefont {Zhao}}, \bibinfo {author} {\bibfnamefont {F.}~\bibnamefont
  {Hu}}, \bibinfo {author} {\bibfnamefont {J.}~\bibnamefont {Sun}}, \ and\
  \bibinfo {author} {\bibfnamefont {B.}~\bibnamefont {Shen}},\ }\href {\doibase
  https://doi.org/10.1016/j.scriptamat.2020.06.034} {\bibfield  {journal}
  {\bibinfo  {journal} {Scripta Materialia}\ }\textbf {\bibinfo {volume}
  {187}},\ \bibinfo {pages} {268} (\bibinfo {year} {2020})}\BibitemShut
  {NoStop}%
\bibitem [{\citenamefont {Soumyanarayanan}\ \emph {et~al.}(2016)\citenamefont
  {Soumyanarayanan}, \citenamefont {Reyren}, \citenamefont {Fert},\ and\
  \citenamefont {Panagopoulos}}]{Soumyanarayanan2016}%
  \BibitemOpen
  \bibfield  {author} {\bibinfo {author} {\bibfnamefont {A.}~\bibnamefont
  {Soumyanarayanan}}, \bibinfo {author} {\bibfnamefont {N.}~\bibnamefont
  {Reyren}}, \bibinfo {author} {\bibfnamefont {A.}~\bibnamefont {Fert}}, \ and\
  \bibinfo {author} {\bibfnamefont {C.}~\bibnamefont {Panagopoulos}},\
  }\href@noop {} {\bibfield  {journal} {\bibinfo  {journal} {Nature}\ }\textbf
  {\bibinfo {volume} {539}},\ \bibinfo {pages} {509} (\bibinfo {year}
  {2016})}\BibitemShut {NoStop}%
\bibitem [{\citenamefont {Battilomo}\ \emph {et~al.}(2021)\citenamefont
  {Battilomo}, \citenamefont {Scopigno},\ and\ \citenamefont
  {Ortix}}]{PhysRevResearch.3.L012006}%
  \BibitemOpen
  \bibfield  {author} {\bibinfo {author} {\bibfnamefont {R.}~\bibnamefont
  {Battilomo}}, \bibinfo {author} {\bibfnamefont {N.}~\bibnamefont {Scopigno}},
  \ and\ \bibinfo {author} {\bibfnamefont {C.}~\bibnamefont {Ortix}},\ }\href
  {\doibase 10.1103/PhysRevResearch.3.L012006} {\bibfield  {journal} {\bibinfo
  {journal} {Phys. Rev. Research}\ }\textbf {\bibinfo {volume} {3}},\ \bibinfo
  {pages} {L012006} (\bibinfo {year} {2021})}\BibitemShut {NoStop}%
\bibitem [{\citenamefont {Bera}\ and\ \citenamefont
  {Mandal}(2021)}]{10.1088/1361-648X/abf783}%
  \BibitemOpen
  \bibfield  {author} {\bibinfo {author} {\bibfnamefont {S.}~\bibnamefont
  {Bera}}\ and\ \bibinfo {author} {\bibfnamefont {S.~S.}\ \bibnamefont
  {Mandal}},\ }\href
  {http://iopscience.iop.org/article/10.1088/1361-648X/abf783} {\bibfield
  {journal} {\bibinfo  {journal} {Journal of Physics: Condensed Matter}\ }
  (\bibinfo {year} {2021})}\BibitemShut {NoStop}%
\bibitem [{\citenamefont {Ho}\ \emph {et~al.}(2019)\citenamefont {Ho},
  \citenamefont {Tan}, \citenamefont {Goolaup}, \citenamefont {Oyarce},
  \citenamefont {Raju}, \citenamefont {Huang}, \citenamefont
  {Soumyanarayanan},\ and\ \citenamefont
  {Panagopoulos}}]{PhysRevApplied.11.024064}%
  \BibitemOpen
  \bibfield  {author} {\bibinfo {author} {\bibfnamefont {P.}~\bibnamefont
  {Ho}}, \bibinfo {author} {\bibfnamefont {A.~K.}\ \bibnamefont {Tan}},
  \bibinfo {author} {\bibfnamefont {S.}~\bibnamefont {Goolaup}}, \bibinfo
  {author} {\bibfnamefont {A.~G.}\ \bibnamefont {Oyarce}}, \bibinfo {author}
  {\bibfnamefont {M.}~\bibnamefont {Raju}}, \bibinfo {author} {\bibfnamefont
  {L.}~\bibnamefont {Huang}}, \bibinfo {author} {\bibfnamefont
  {A.}~\bibnamefont {Soumyanarayanan}}, \ and\ \bibinfo {author} {\bibfnamefont
  {C.}~\bibnamefont {Panagopoulos}},\ }\href {\doibase
  10.1103/PhysRevApplied.11.024064} {\bibfield  {journal} {\bibinfo  {journal}
  {Phys. Rev. Applied}\ }\textbf {\bibinfo {volume} {11}},\ \bibinfo {pages}
  {024064} (\bibinfo {year} {2019})}\BibitemShut {NoStop}%
\bibitem [{\citenamefont {Tan}\ \emph {et~al.}(2021)\citenamefont {Tan},
  \citenamefont {Liu},\ and\ \citenamefont {Yan}}]{PhysRevB.103.214438}%
  \BibitemOpen
  \bibfield  {author} {\bibinfo {author} {\bibfnamefont {H.}~\bibnamefont
  {Tan}}, \bibinfo {author} {\bibfnamefont {Y.}~\bibnamefont {Liu}}, \ and\
  \bibinfo {author} {\bibfnamefont {B.}~\bibnamefont {Yan}},\ }\href {\doibase
  10.1103/PhysRevB.103.214438} {\bibfield  {journal} {\bibinfo  {journal}
  {Phys. Rev. B}\ }\textbf {\bibinfo {volume} {103}},\ \bibinfo {pages}
  {214438} (\bibinfo {year} {2021})}\BibitemShut {NoStop}%
\bibitem [{\citenamefont {Mohanta}\ \emph {et~al.}(2020)\citenamefont
  {Mohanta}, \citenamefont {Okamoto},\ and\ \citenamefont
  {Dagotto}}]{PhysRevB.102.064430}%
  \BibitemOpen
  \bibfield  {author} {\bibinfo {author} {\bibfnamefont {N.}~\bibnamefont
  {Mohanta}}, \bibinfo {author} {\bibfnamefont {S.}~\bibnamefont {Okamoto}}, \
  and\ \bibinfo {author} {\bibfnamefont {E.}~\bibnamefont {Dagotto}},\ }\href
  {\doibase 10.1103/PhysRevB.102.064430} {\bibfield  {journal} {\bibinfo
  {journal} {Phys. Rev. B}\ }\textbf {\bibinfo {volume} {102}},\ \bibinfo
  {pages} {064430} (\bibinfo {year} {2020})}\BibitemShut {NoStop}%
\end{thebibliography}
\end{document}